\documentclass[aps,prd,showpacs,reprint,floatfix,preprintnumbers,nofootinbib]{revtex4-1}
\usepackage{amsmath,latexsym,amssymb,amsfonts,epsfig,xspace,amsthm}
  
\numberwithin{equation}{section}
\allowdisplaybreaks

\newcommand{\R}{\mathbb{R}}                  
\newcommand{\Z}{\mathbb{Z}}                  

\newcommand{\lqg}{loop quantum gravity\xspace}

\newcommand{\oiint}{\bigcirc \hspace{-1.35em} \int \hspace {-0.8em} \int}
\newcommand{\scpr}[2]{\langle #1 \vert  #2 \rangle}
\newcommand{\sscpr}[3]{\langle #1 \vert #2 \vert #3 \rangle}

\newcommand{\ket}[1]{\rvert #1 \rangle}
\newcommand{\norm}[1]{\left\lVert #1 \right\rVert}

\newcommand{\poi}[2]{\left\{#1 , #2 \right\}}
\newcommand{\expec}[1]{\langle #1 \rangle}

\newcommand{\dmal}{\text{d}\mu_{\text{AL}}}

\newcommand{\ppexp}[1]{\mathcal{P}\exp \oiint_{#1}}

\newcommand{\pb}[1]{\underset{\Leftarrow}{#1}}

\newcommand{\ch}[1]{\underline{#1}}

\newcommand{\lp}{\ell_\text{p}}
\newcommand{\spat}{\mathcal{S}}
\newcommand{\op}[1]{\boldsymbol{#1}}
\newcommand{\dd}{\text{d}}

\DeclareMathOperator{\abar}{\overline{\mathcal{A}}}

\DeclareMathOperator{\tr}{tr}

\DeclareMathOperator{\um}{\mathbb{I}}

\DeclareMathOperator{\alg}{\mathfrak{A}}
\DeclareMathOperator{\cyl}{Cyl}

\begin{document}
\date{23 Apr 2011}
\title{Black hole horizons from within loop quantum gravity}
\author{Hanno Sahlmann}
\email{sahlmann@apctp.org}
\affiliation{Asia Pacific Center for Theoretical Physics, Pohang (Korea)}
\affiliation{Physics Department, Pohang University for Science and Technology, Pohang (Korea)}
\preprint{APCTP Pre2011-004}

\begin{abstract}
In general relativity, the fields on a black hole horizon are obtained from those in the bulk by pullback and restriction. Similarly, in quantum gravity, the quantized horizon degrees of freedom should result from restricting, or pulling-back, the quantized bulk degrees of freedom. 
This is not yet fully realized in the -- otherwise very successful --  quantization of isolated horizons in loop quantum gravity. In this work we outline a setting in which the quantum horizon degrees of freedom are simply components of the quantized bulk degrees of freedom. There is no need to quantize them separately. We present evidence that for a horizon of sphere topology, the resulting horizon theory is remarkably similar to what has been found before.  
\end{abstract}
\maketitle
\section{Introduction}
The quantization of an isolated horizon is a remarkable success of \lqg \cite{Smolin:1995vq,Rovelli:1996dv,Ashtekar:1997yu,Ashtekar:2000eq,Kaul:1998xv,Kaul:2000kf,Domagala:2004jt,Meissner:2004ju,Corichi:2006wn,Engle:2009vc,Engle:2010kt,Agullo:2010zz,Engle:2011vf}. However, it is only an effective description, in the sense that it uses a number of elements that are not intrinsic to the formalism of loop quantum gravity. For example, the location of the horizon is fixed to be the boundary of the space-time, and the fields on the boundary, although related to those in the bulk, are quantized separately, using a symplectic structure that is derived from the one on the bulk fields in the classical theory \cite{Ashtekar:2000eq,Engle:2010kt}. 
 
The goal of the present work is to advocate a slightly more intrinsic viewpoint. For this, we take as the input from the classical theory \emph{only} the horizon boundary conditions 
\begin{equation}
\pb{F}(A)=-\frac{\pi(1-\beta^2)}{a_H}\pb{\Sigma}(E).
\label{eq:horizon}
\end{equation}
Here, $A$ and $E$ are the canonical variables of \lqg, $F$ is the curvature of $A$, $\Sigma$ is the dual of $E$, and the arrows denote pullback to a given surface $H$ that is an isolated horizon \cite{Ashtekar:1998sp} of type I. We have stated here the SU(2) isolated horizon condition from \cite{Engle:2009vc,Engle:2010kt} since it uses less external input. We will however also be treating a model with the U(1) condition from  \cite{Ashtekar:1997yu,Ashtekar:2000eq}.
States $\Psi$ that contain a black hole horizon are then solutions to an equation -- in the quantum theory -- of the structure 
\begin{equation}
\widehat{\pb{F}}\,\Psi=-\frac{\pi(1-\beta^2)}{a_H}\widehat{\pb{\Sigma}}\,\Psi
\label{eq:q_horizon}
\end{equation}
where the operators on both sides are defined in terms of elementary operators of \lqg.  We will also refer to surfaces $S$ in \eqref{eq:q_horizon} as \emph{horizon branes}. They can be thought of as loci of very highly excited quantum fields -- such that \eqref{eq:q_horizon} is satisfied. Such branes are not to be found in the kinematic Hilbert space of \lqg. Rather, these states lie in different representations of the holonomy-flux algebra, due to the branelike excitations. 

In fact, the present work can be viewed as a continuation of some lines of thought in the earliest work \cite{Smolin:1995vq} connecting horizons in loop quantum gravity to topological quantum field theory. \cite{Smolin:1995vq} was prescient in many ways, for example by introducing SU(2) boundary conditions similar to \eqref{eq:horizon}, for identifying Chern-Simons theory as describing the horizon degrees of freedom, and for linking the size of the horizon state space to the entropy of a black hole. It also already contained the idea that the observables on the horizon should form a subablgebra of the full algebra of gravity observables. 

What we will do is to collect evidence that the condition \eqref{eq:q_horizon} allows for solutions $\Psi$ that are remarkably close in  structure to what has been found upon quantizing the Chern-Simons phase space on the horizon. In particular, we will present evidence that, by \emph{restricting} such states and the \lqg operators to the horizon, one will obtain a theory that resembles quantum SU(2) BF theory in the spherical case, or, for general horizon topologies, ISU(2) Chern-Simons theory \cite{Noui:2004iy,Noui:2004iz}.
Thus, for the case of spherical topology, which is most relevant for the description of black holes, we seem to obtain a very similar state counting. Thus, no separate quantization of the horizon degrees of freedom seems to be necessary. Those degrees of freedom are already part of the quantized gravitational field of \lqg. 

Let us sketch the evidence that we have: For one thing, for a model in which the structure group SU(2) is replaced by U(1), we can find exact solutions to the analog of condition \eqref{eq:q_horizon}. The resulting surface theory resembles U(1) BF-theory coupled to particles. 
For the case of SU(2), we do not have all the technical details in hand. Our preliminary analysis shows, however, that a gauge invariant state that solves \eqref{eq:q_horizon} is, when restricted to the brane, a solution of the constraints of SU(2) BF theory, at least formally.

Obviously, our proposal, even if made fully rigorous in the SU(2) case, is not a fully quantum-mechanical description of black holes. For example, the horizon area still appears in \eqref{eq:q_horizon}. It is, rather, one step in this direction. In fact, more radical proposals have been made \cite{Rovelli:1996dv,Krasnov:1998vc,Livine:2005mw,Krasnov:2009pd}. 

Some of the ideas and results contained in this work have already been implicitly or explicitly articulated in the literature on quantum isolated horizons. We have already mentioned \cite{Smolin:1995vq}. Reference \cite{Ashtekar:2000eq} contains a detailed discussion of how to split the space of generalized connections into a boundary and a bulk part, and how the connections appearing in U(1) Chern-Simons theory with particles define generalized connections. 
As another example, in \cite{Engle:2010kt} certain operators in SU(2) Chern-Simons theory are \emph{identified} with certain \lqg operators. What is new in the present work is that we take these ideas as far as possible. 

In the next section, we will use heuristic considerations to support the new picture. In Sec.\ \ref{se:u1} we discuss, in some technical detail, a U(1) model. Sec.\ \ref{se:su2} contains the results we obtained for the SU(2) case. We finish with a discussion of the results and open questions in Sec.\ \ref{se:disc}.  
\section{Heuristic considerations}
\label{se:heu}
In general relativity, the fields on a black hole horizon are obtained from those in the bulk by pullback or restriction. Similarly, 
in quantum gravity, the quantized horizon degrees of freedom should result from restricting, or pulling-back, in a suitable way the quantized bulk degrees of freedom. In the previous literature on quantum isolated horizons, the pullbacks of the bulk degrees of freedom have been quantized separately, starting from a symplectic structure that was obtained from a boundary term in the symplectic structure of general relativity. Here we want to proceed differently: We start from the holonomy-flux algebra $\mathfrak{A}$ which is a quantization of the kinematic degrees of freedom of general relativity in the connection formulation. This algebra makes no reference to horizons or branes whatsoever. Then we will look for representations of $\mathfrak{A}$ which contain states that solve \eqref{eq:q_horizon}. Finally, once we have solutions of \eqref{eq:q_horizon} in hand, we can consider the action on these solutions, by operators  localized, in a suitable sense, in $H$. This  constitutes the `horizon theory'.  

Let us start by considering the classical theory, and make an inventory of the degrees of freedom on the horizon. In the canonical formulation, the gravitational fields live on a spatial slice $\spat$ of space-time. We take $\spat$ to be orientable and oriented. In terms of connection variables, the fields are $A$ (an su(2) valued connection one form) and $E$ (a triad of vector densities). They are coordinates in a phase space given by the Poisson brackets
\begin{equation}
\poi{A^i_a(x)}{E^b_j(y)}=8\pi G \beta \delta_a^b\delta_i^j\delta(x,y).  
\label{eq:poi}
\end{equation}  
We now consider a two dimensional submanifold $H$ of $\spat$. We note that with a view towards black hole horizons, the case of $H$ being homeomorphic to $S^2$ is the most relevant one. At this point, we will however only assume that $H$ is compact, connected, and orientable, and we will chose an orientation. Note also that $H$ need not be a boundary of $\spat$. If $\spat$ has a boundary, then $H$ may or may not be part of that boundary.  

If we restrict attention to $H$, we can divide up $A$ and $E$ as follows. First, we have the components intrinsic to $H$, the pullbacks
\begin{equation}
\pb{E}:= i_H^*E, \qquad    \pb{A}:= i_H^*A
\end{equation}
where $i_H$ is the embedding of $H$ in $\spat$.  We note that these fields precisely correspond to the kinematic canonical variables of SU(2) BF theory, if the brackets \eqref{eq:poi} are extended in a suitable way to the pullbacks.

Next, there are the remaining components of $A$ and $E$ on $H$. These can be given as 
\begin{equation}
A_\perp:= A(n), \qquad  \pb{\Sigma}:= i_H^* \Sigma
\end{equation}
where $n$ is a fixed transversal vector field  on $H$, and $\Sigma$ is the dual of $E$, 
\begin{equation}
\Sigma_i
=\epsilon_{ijk}e^j\wedge e^k
=\frac{1}{2}E^a_i\epsilon_{abc}\,dx^b\wedge dx^c,
\end{equation}
with $e$ the spatial triad. The conditions \eqref{eq:horizon} link $\pb{A}$ with $\pb{\Sigma}$. Fixing $\Sigma$ and imposing the conditions on $H$ completely fixes the curvature of  
$\pb{A}$ and hence most of the gauge invariant degrees of freedom contained in 
$\pb{A}$. What remains are essentially the holonomies around nontrivial cycles in $H$. Again we note that this is in analogy to SU(2) BF theory, now after imposition of the constraints. 

After having organized the classical degrees of freedom on the horizon, we now come to the quantum theory. Here the object we have to consider is the \emph{holonomy-flux} algebra $\alg$ \cite{Ashtekar:1994wa,Ashtekar:1994mh,Sahlmann:2002xv,Lewandowski:2005jk}, since it encodes, on an abstract level, the quantization chosen in \lqg. 
The elements of $\alg$ corresponding to $A$ can be thought of as functions of the holonomies of $A$. They form an Abelian subalgebra $\cyl$. The elements corresponding to the densitized triad $E$ are quantizations of the `fluxes'
\begin{equation}
E_{S,f}= 2\int_S f^I\Sigma_I, 
\end{equation} 
where $S$ is a surface in $\spat$. The pullback of $A$ on $H$ is then encoded in the cylindrical functions that just depend on holonomies in $H$.  

The pullback of $E$ is encoded in the flux through $S\cap H$ for surfaces $S$ transversal to $H$. Operators corresponding to such `fluxes through one-dimensional submanifolds' are sometimes also considered part of $\mathfrak{A}$ (as in \cite{Sahlmann:2003in,Fleischhack:2004jc}) and can, in any case, be defined in the Ashtekar-Lewandowski (AL) representation of $\mathfrak{A}$. Even if one does not want to consider these operators, the information about the pullback of $\Sigma$ is certainly contained in the quantized flux through $S$ for surfaces $S$ transversal to $H$.

The component $A_\perp$ is quantized in holonomies transversal to $H$, and the pullback of $\Sigma$ has its exclusive quantization by flux operators for surfaces $S$ \emph{within} $H$. 

Now that we have accounted for the degrees of freedom on $H$ in $\mathfrak{A}$, we can try to answer the question: Can we consistently ask for condition \eqref{eq:q_horizon}, given the commutation relations in $\mathfrak{A}$? 

\begin{figure}
	\centerline{\epsfig{file=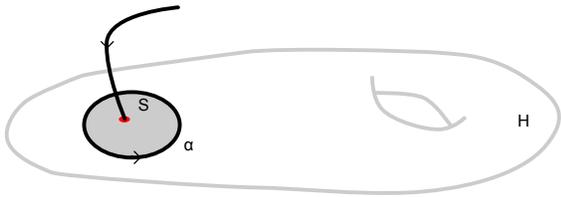,scale=.6}}
	\caption{A spin network edge punctures $H$ that results in nontrivial holonomy around $\partial S= \alpha$. }
	\label{fi:flux_hol}
\end{figure}

The first thing to note is that 
\begin{equation}
[E_{S,f}, h_\alpha]=0 \quad \text{ for }\quad  \alpha, S \subset H, 
\label{eq:comm}
\end{equation}
which is good news as \eqref{eq:q_horizon} would become extremely restrictive, if not inconsistent, if the pullbacks of $\Sigma$ and $A$ were not commuting.   
The next thing to note is that only holonomies are contained in $\mathfrak{A}$, not the connection $A$ itself, nor its curvature. Thus \eqref{eq:q_horizon} can not be imposed as it stands. Fortunately, the 
non-Abelian Stokes Theorem (see for example \cite{nastokes}) relates surface integrals of curvature with holonomy, 
\begin{equation}
h_\alpha[A]=\ppexp{S} \mathcal{F}[A].
\label{eq:stokes}
\end{equation}
Here $\alpha$ is a loop that bounds the surface $S$, $\mathcal{F}=hFh^{-1}[A]$ is the curvature $F=D_A A$ of
$A$, transported to the beginning/endpoint of $\alpha$, and the surface integral on the right-hand side is \textit{surface ordered}. We should mention that for this formula to hold, $S$ must be simply connected. Using Stokes' Theorem, one can thus replace certain functionals depending on the curvature, with functionals depending on the connection, in the classical theory. The idea is, then, to replace \eqref{eq:q_horizon} with 
\begin{equation}
h_{\partial S}\,\Psi=
\ppexp{S}-\frac{2\pi(1-\beta^2)}{a_H}h\Sigma h^{-1}\,\Psi. 
\label{eq:exph}
\end{equation}
Since surface integrals of $\Sigma$ act nontrivially only at transverse intersections by holonomies, a spin network edge which punctures $H$ will correspond to nontrivial holonomy around $\partial S=\alpha$, see Fig.\ \ref{fi:flux_hol}. 
But now an immediate concern is whether it is possible to define the complicated operator -- let us call it $\op{W}_S$ -- on the right-hand side. For the \emph{trace} of $\op{W}_S$ in the $j=1/2$-representation this question was answered affirmatively in \cite{Sahlmann:2011uh}. We will describe some of the details below, in Sec.\ \ref{se:su2}. Here it suffices to say that the main difficulty in defining the right-hand side is that because the components of $\Sigma$ do not commute, there is an ordering ambiguity. The authors of \cite{Sahlmann:2011uh} pointed out a way to resolve this ordering ambiguity by using a device from the theory of Lie algebras, the \emph{Duflo map}. Using this ordering, spin network functions are eigenstates of the operator $\tr\op{W}_S$ under many circumstances, and the corresponding eigenvalues are related to path integral expectation values of Wilson loops in SU(2) Chern-Simons theory. 

We remark that there is an apparent contradiction between the fact, that holonomies in $\mathfrak{A}$ commute, whereas the operators $\op{W}_S$ certainly do not commute with holonomies in general. So how can it be equal to a holonomy? The resolution is that it is certainly not equal to a holonomy in general, \eqref{eq:exph} being a highly nontrivial condition on the state. The only thing that follows from this consideration is that the action of holonomy operators on a solution $\Psi$ of \eqref{eq:exph} can, in general, not be a solution again.

While both sides of \eqref{eq:exph} are defined in the AL representation, the standard representation of \lqg, there are no solutions to \eqref{eq:exph} in this representation. To find solutions of \eqref{eq:exph}, there are then at least two possible strategies: One can take the properties of the operators on the left-hand side in the AL representation of $\mathfrak{A}$, and use them to define a nonstandard representation of the operators on the right-hand side, or, vice versa, use the standard representation of the right to seek a nonstandard representation of the left-hand side. We chose the latter possibility in this article, for the reason that the results seem to compare well with previous work. 

In the AL representation of $\mathfrak{A}$, the action of flux operators $E_{S,f}$ is concentrated on transversal intersections of holonomies with $S$, and this continues to hold for the operator $\tr\op{W}_S$. Thus we need to find a state such that this is the case also for holonomies within $H$. We can use the fact that the space of  generalized connections factorizes into a space of connections on $H$, and a space of connections in the `bulk'  
\begin{equation}
\abar = \abar_H\times\abar_\perp. 
\end{equation}
The standard representation is defined by the AL measure on $\abar$. 
The idea is now to modify this measure on $\abar_H$ in a suitable way.  Essentially what one wants to define are measures  
\begin{equation}
\delta^{(\abar_H)}\left(F[A](x)-\sum_i c_i\delta^{(H)}(p_i,x)\right) \dmal\rvert_{\abar_H}\times \dmal\rvert_{\abar_\perp}
\label{eq:meas}
\end{equation}
where the first delta function is a \emph{functional} Dirac-delta-function, and the second one is the ordinary Dirac delta  on $H$. In the Hilbert space generated by such a measure, one can then find solutions to \eqref{eq:exph}, by considering spin networks that end in the points $p_i$, and adjusting the constants $c_i$ appropriately. In the U(1) model, this construction  goes through quite literally. In the SU(2) case, we have no rigorous proof that, with the right-hand side evaluated on a state in the AL representation, \eqref{eq:exph} rigorously defines a measure on $\abar_H$, of the form indicated above. But it is clear that, with the right-hand side well defined, \emph{the holonomies on all contractible loops in $H$ are fixed}, at least up to conjugation, and thus there are no local gauge invariant degrees of freedom left in the holonomies on $H$. There \emph{are}, however, holonomies that are not fixed by \eqref{eq:exph}, holonomies that run between punctures, and also around nontrivial cycles of $H$ if not simply connected. There are thus  nontrivial holonomy operators for those, and thus provided that the measures can be constructed rigorously, all the observables related to the connection are represented.

What about the fluxes? What happens to the flux operators when one changes the measure has been studied in detail in \cite{Sahlmann:2002xu,Sahlmann:2002xv}. The upshot is that if one modifies the measure, one has to modify the fluxes by adding a divergence term, 
\begin{equation}
\pi(E_{S,f})=X_{S,f}+ \frac{i}{2}\text{div}_\mu(S,f) 
\label{eq:div}
\end{equation}
to have the operators still symmetric. Here $X_{S,f}$ is a certain derivation on the cylindrical functions. This divergence can be ill-defined (more precisely, not $L^2$ as would be required) for a 
delta-function in the measure, thus existence of symmetric flux operators $E_S$ for which $S$ touches the surface is, at first sight, questionable. But it turns out that all the fluxes one needs are actually well defined for a measure of the form \eqref{eq:meas}. First off, the delta function in \eqref{eq:meas} concerns only $\pb{A}$, but $\pb{\Sigma}$ corresponds to an operator acting on $A_\perp$. Thus all the fluxes $E_S$ with $S\subset H$ are well defined and symmetric without any change. Moreover, whenever a holonomy variable is constrained by the delta function in the measure \eqref{eq:meas}, there is \emph{no} corresponding degree of freedom in the quantum theory, hence no way for any fluxes to act in a nontrivial way. Vice versa, the action of the flux operators is well defined on holonomies that represent degrees of freedom  leftover under the measure. To summarize, some flux operators may not be well defined, but their failure to exist can be understood easily, and they are not needed for the physical interpretation of the resulting theory, anyway.  

Let us make some remarks about the properties of the measure \eqref{eq:meas} and the consequences for the state spaces: The first one is that in the case of $H$ being a topological sphere, there are no nontrivial cycles, so the only degrees of freedom on $H$ reside in holonomies connecting the punctures (and in the conjugate fluxes). Gauge invariant states are constructed out of such holonomies, and holonomies in the bulk, by forming spin networks. A priori, there are many different ways of forming the spin network component on $H$ but many of them will describe the same state, due to the flatness constraint built into the measure. In fact we will argue that the independent states on $H$, given the punctures, are labeled by a \emph{single intertwiner} between the spins at the punctures. This is very close to the pictures in \cite{Ashtekar:2000eq,Engle:2009vc}. Note that this is fully born out in the U(1) case.      

The second remark is that, again due to the flatness constraint built into the measure, the exchange of punctures leads to the same state up to a nontrivial phase. Thus the punctures enjoy a nonstandard statistics, and counting states modulo diffeomorphisms is nontrivial.  This is again reminiscent of \cite{Ashtekar:2000eq,Engle:2009vc}, and it is actually vital to obtain proportionality between area and entropy of the horizon. 

The arguments that we have given so far are somewhat heuristic, if encouraging. We will show, however, that they can be made 
completely precise in a model with structure group U(1), to which we turn next. We will then begin to address the  
case of structure group SU(2). We will discuss more details in Sec.\ \ref{se:su2}. 

\section{U(1) boundary conditions}
\label{se:u1}
In this section, we consider the kinematics of \lqg, but with the structure group SU(2) replaced by U(1) \cite{Corichi:1997us}. 
In this model we replace condition \eqref{eq:q_horizon} by 
\begin{equation}
h_{\partial S} \Psi= e^{-\frac{2\pi i\beta}{a_H} E_S}\Psi, 
\label{eq:hu1}
\end{equation}
between a quantized U(1) connection $A$ and a quantized vector density $E$. 
Note that this is precisely the isolated horizon condition obtained by gauge fixing to U(1) as used in \cite{Ashtekar:2000eq}. Much of the material of this section will remain valid if the gauge fixing is just carried out on the horizon, in particular the quantum theory on the horizon as far as the connection is concerned. 

Irreducible representations of U(1) are labeled by integers, and hence the generalized spin networks correspond to functions 
\begin{equation}
T_{\gamma,\ch{n}}[A]=\prod_{e\in \gamma} (h_e[A])^{n_e},    
\end{equation}
which are usually called charge networks. They are gauge invariant, whenever the incoming charges equal the outgoing charges at each vertex, 
\begin{equation}
\sum_{e \text{ into } v}n_e= \sum_{e \text{ out of } v} n_e \quad\text{ for all }v.
\end{equation}
Charge nets commute with the operator $E_S$ as follows:
\begin{equation}
\begin{split}
[E_S, T_{\gamma,\ch{n}}] &= X_S[T_{\gamma,\ch{n}}]\\
&= 2\pi\beta \lp^2 \left[\sum_{v\in\gamma\cup S}\, \sum_{e \text{ at } v} \sigma(e) n_e\right] \, T_{\gamma,\ch{n}},
\label{eq:su1}
\end{split}
\end{equation}
where it is assumed that all edges intersect $S$ in vertices of $\gamma$ and $\sigma$ is +1, -1, or 0, depending on whether the edge is oriented  consistently\footnote{What we mean by `consistent' is the following:
Both $S$ and $\spat$ carry orientations. Let $(s_1,s_2)$ be a positively oriented basis of tangent vectors to $S$. Then if $(s_1,s_2,t)$ is positively oriented in $\spat$, with $t$ the tangent of $e$ in the intersection point, then we call the orientations of $e$ and $S$ consistent.}
with the surface, the opposite way as the surface, or is tangential to the surface. $X_S$ is a derivation on the space of charge nets. 

Given a charge network, we can always decompose it as 
\begin{equation}
T_{\gamma,\ch{n}}=T_{\gamma^H,\ch{n}^H} T_{\gamma^\perp,\ch{n}^\perp}
\label{eq:decomp}
\end{equation}
where $\gamma^H\subset H$ and $\gamma^\perp$ intersects $H$ only transversally. 
Thus given $(\gamma,\ch{n})$, in view of \eqref{eq:hu1},\eqref{eq:su1} we need a state $\Psi$ such that for loops $\partial S$ in $H$, we would have 
\begin{equation}
h_{\partial S}\,\Psi=
e^{-\frac{2\pi i}{k}
\sum\limits_{p\in S\cap\gamma^\perp} m_p}\Psi,
\label{eq:horu1}
\end{equation}
where we have set $k=a_H/\lp^2$ and $m_p=\sum_{e \text{ at } p} \sigma(e) n_e$. We will call 
\begin{equation}
\mathcal{P} = \{(p_1,m_1),(p_2,m_2), \ldots, (p_N,m_N)\}
\end{equation}
puncture data. Equation \eqref{eq:horu1} means that the connection on $H$ must be a (quantized) flat connection,
\begin{equation}
F(A)(x)= -2\pi/k \sum_i m_i\delta(x-p_i),
\label{eq:flat}
\end{equation}
We will see however, that due to the fact that \eqref{eq:horu1} only speaks about holonomies, not the curvature itself, if we change the puncture data, by adding to each $m_i$ a multiple of $k$, the quantum state on the horizon will not change. Thus one may also view the $m_i$ as elements of $\Z_k$. 
\subsection{Lebesgue measure on flat connections}
We will now define a functional on charge networks that can be regarded as Lebesgue integral on connections that fulfill \eqref{eq:flat} for some puncture data $\{(p_1,m_1),(p_2,m_2), \ldots, (p_N,m_N)\}$. We will see that in the case that $H\simeq S^2$ the functional is only well defined if 
\begin{equation}
\sum_i m_i = 0 \mod k,
\label{eq:zeromodk}
\end{equation}
and that the actual parameters of the state we construct are not the $m_i$ but the $m_i \mod k$. Let us also define $H'=H-\{p_1,\ldots p_N\}$. 

It is clear what to do in principle: A given charge network $T_{\gamma,\ch{n}}$ with $\gamma\subset H$ needs to be decomposed into factors such that we can apply \eqref{eq:horu1} to split the variables into ones that are free, and ones that are determined. But this has to happen in a consistent way, and the bookkeeping involved in doing this by hand gets unwieldy very quickly. Fortunately homology theory comes to the rescue (see for example 
\cite{frankel} for a gentle introduction to the concepts used below). Note first that by subdividing and adding edges, $\gamma$ can always be made into the 1-skeleton of a subcomplex of the singular chain complex of $H'$. Then, given this subcomplex, a labeling of the graph edges with charges $\ch{n}$ defines a 1-chain
\begin{equation}
\ch{n}=\sum{n_i}{e_i}
\end{equation}
with integer coefficients.      
$T_{\gamma,\ch{n}}$ is gauge invariant precisely when $\ch{n}$ is a cycle, $\partial\ch{n}=0$. There is a natural pairing between chains and one-forms, and in particular between the chain $\ch{n}$ and connections $A$, 
\begin{equation}
\scpr{\ch{n}}{A}=\sum_i n_i\int_{e_i} A.  
\end{equation}
The connections $A$ relevant for the definition of the functional are flat, $\dd A=0$. Let us assume for the moment that $\partial\ch{n}=0$. Then the above pairing is gauge invariant, and $\scpr{\ch{n}}{\cdot}$ is a functional on the first de Rham cohomology $H^1(H';\R)$. Let $\{\ch{l}_i\}$ be a basis of elementary cycles of $H'$, and let $\{a_i\}$ be the dual basis in $H^1(H';\R)$. Then we introduce parameters $\phi_i$ to write 
\begin{equation}
a(\phi)=\sum_i \phi_i a_i. 
\end{equation}
Now we can define the state. Let 
\begin{equation}
\begin{split}
\mu'(T_{\gamma,\ch{n}})
=\int_0^{2\pi}\ldots & \int_0^{2\pi}\\ 
&e^{\scpr{\ch{n}}{a(\phi)}}
\prod_{p_i} 2\pi \delta(\phi_i+2\pi n_i/k)
\prod_j \frac{\text{d}\phi_j}{2\pi}.
\end{split}
\label{eq:state}
\end{equation}
Note that this formula is manifestly invariant under subdivison of the graph underlying $\ch{n}$, and under adding new edges. It is thus \emph{consistent}
with the equivalence of labelings of charge network functions.
It also defines a positive functional. To see this, let $f=\sum_I c_I T_{\gamma,\ch{n}_I}$. Then
\begin{equation}
\begin{split}
\mu'(|f|^2)&=\sum_{IJ}\overline{c_I}c_J \int e^{\scpr{\ch{n}_J-\ch{n}_I}{a(\phi)}}\prod_{p_i} \delta(\ldots) \text{d}\phi\\
&=\int \left|\sum_Ic_I e^{\scpr{\ch{n}_I}{a(\phi)}}\right|^2
\prod_{p_i} \delta(\ldots) \text{d}\phi\\
&\geq 0.
\end{split}
\end{equation}
Before we study further properties, let us extend this definition to 
a state $\mu$ on not necessarily gauge invariant charge networks, by simply declaring $\mu$ to be $\mu'$ on gauge invariant charge networks, and zero otherwise. More formally, let 
\begin{equation}
\mu(T_{\gamma,\ch{n}}):=\mu''(T_{\gamma,\ch{n}})\mu'(T_{\gamma,\ch{n}})
\text{ with } \mu''(T_{\gamma,\ch{n}})=\delta(\partial\ch{n}).
\end{equation} 
This does not spoil positivity: 
It can be easily seen that $\mu''$ is positive. Thus, given charge nets $T_{\gamma,\ch{n}_I}$, the matrices
\begin{equation}
M'_{IJ}=\mu'(T_{\ch{n}_J-\ch{n}_I}), \qquad M''_{IJ}=\mu''(T_{\ch{n}_J-\ch{n}_I})
\end{equation}
are  positive semidefinite. But then the Hadamard product $M' \circ M''$ (obtained by multiplying the matrices entrywise) is positive semidefinite according to the Schur Product Theorem, and hence for  $f=\sum_I c_I T_{\gamma,\ch{n}_I}$
\begin{equation}
\mu(|f|^2)=\sum_{IJ} \overline{c}_{I}c_{J} (M''\circ M')_{IJ}\geq 0.   
\end{equation}
Now we discuss the GNS representation given by $\mu$. We will denote the ground state by $\ket{0}_{\mathcal{P}}$, and the GNS state corresponding to $T_{\gamma,\ch{n}}$ by $\ket{\ch{n}}_{\mathcal{P}}$. The dependence on $\gamma$ is left implicit to improve readability. Note that since $\mu$ has a large kernel, many GNS vectors actually have zero norm. 

First, let $\alpha$ be the boundary of a surface $S$ in $H$. Then either $\alpha$ is contractible, in which case for the corresponding cycle we can write $\partial S=\alpha$, and note 
\begin{equation}
\scpr{\alpha}{a(\phi)}=\scpr{\partial S}{a(\phi)}=\scpr{S}{\dd a(\phi)}=0.
\end{equation}
Or $\alpha$ goes around a puncture, in which case  
$\scpr{\alpha}{a(\phi)}=\phi_j$ for some $\phi_j$ that is in a delta-function
in \eqref{eq:state}. Thus for arbitrary $\ch{n}$ we find 
\begin{equation}
\begin{split}
\sscpr{\ch{n}}{h_\alpha}{0}_{\mathcal{P}}
&=\int e^{i\scpr{\alpha-\ch{n}}{a(\phi)}}\, \text{d}\phi\\
&=\scpr{\ch{n}}{0}_{\mathcal{P}}\cdot
\begin{cases} 1 &\text{ if } \alpha=\partial S\\ 
\exp(-2\pi i n_j/k) & \text{ if $\alpha$ around $p_j$} 
\end{cases}. 
\end{split}
\end{equation}
Since the vectors $\ket{\ch{n}}$ are dense by construction, this shows that $\ket{0}$ is an eigenstate to $h_\alpha$, and that it solves \eqref{eq:horu1}.
Since charge networks commute, the calculation also shows that 
$T_{\gamma,\ch{n}}\ket{0}$ is again a solution to \eqref{eq:horu1}, where $\gamma$ is any graph in $H$. 

\begin{figure}
	\centerline{\epsfig{file=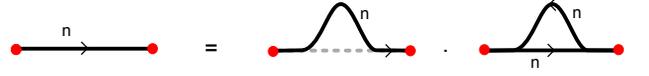,scale=.6}}
	\caption{Relation between a charge network edge and its transformation under a diffeomorphism}
	\label{fi:diffeo1}
\end{figure}

\begin{figure}
	\centerline{\epsfig{file=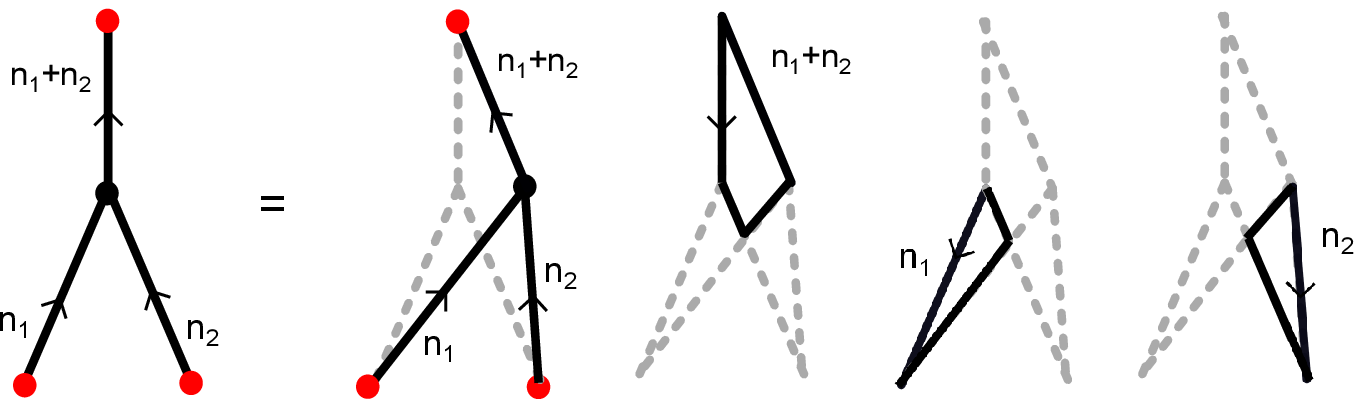,scale=.6}}
	\caption{Relation between a charge network and its tranformation under a diffeomorphism}
	\label{fi:diffeo2}
\end{figure}

Let us now briefly discuss the properties of the state under diffeomorphisms. It turns out that for a diffeomorphism $\varphi$ of $H$ that is connected to the identity and that fixes the punctures, as well as the boundary 0-cycle $\partial \ch{n}$ of a charge net 
$T_{\gamma,\ch{n}}$, one finds   
\begin{equation}
\ket{\gamma,\ch{n}}_{\mathcal{P}}=\ket{\varphi(\gamma),\ch{n}}_{\mathcal{P}}.
\end{equation}
We will not prove this in detail, but just sketch the idea.
Consider first an edge $e$ in $H$ and its image $\varphi(e)$ under a diffeomorphism $\varphi$ that is the identity outside of a compact region, and drags part of $e$ as in Fig.\ \ref{fi:diffeo1}. In this case $T_{e,n}$ and $T_{\varphi(e),n}$ are related by a cycle $\alpha$, 
\begin{equation}
T_{e,n}T_{\varphi(e),-n}=T_{\alpha,n}.
\end{equation}
If $\varphi$ fixes the punctures, $\alpha$ can not contain a puncture, and we find 
\begin{equation}
\begin{split}
\norm{\ket{e,n}-\ket{\varphi(e),n}}_\mathcal{P}^2
&=\omega(|T_{e,n}|^2+|T_{e,n}|^2\\
&\qquad-T_{e,-n}T_{\varphi(e),n}
-T_{e,n}T_{\varphi(e),-n})\\
&=\omega(1+1-T_{\alpha,-n}-T_{\alpha,n})\\
&=0,
\end{split} 
\end{equation}
since $\alpha=\partial S$ and hence $\omega(T_{\alpha})=0$. Thus  
$\ket{e,n}=\ket{\varphi(e),n}$. With a similar calculation, one can show that one can move gauge invariant vertices in a charge network without changing the corresponding GNS state (see Fig.\ \ref{fi:diffeo2}).  
\subsection{Extension to $\mathfrak{A}$}
\label{se:ext}
We can now extend the state we found in the preceding section even further, to charge networks $T_{\gamma,\ch{n}}$ that have graphs in $\spat$. To that end we use the decomposition \eqref{eq:decomp} of a general charge net $T_{\gamma,\ch{n}}$ into the product $T_{\gamma^H,\ch{n}^H} T_{\gamma^\perp,\ch{n}^\perp}$ where the first factor is entirely in $H$ and the second is transversal to $H$. Then set 
\begin{equation}
\omega(T_{\gamma,\ch{n}})=\mu(T_{\gamma^H,\ch{n}^H})\omega_{\text{AIL}}(T_{\gamma^\perp,\ch{n}^\perp})
\end{equation}
where $\omega_{\text{AIL}}$ is the state devised by Ashtekar, Isham and Lewandowski. 
$\omega$ is positive by virtue of the Schur Product Theorem: For $f=\sum_I c_I T_{\gamma,\ch{n}_I}$
we have 
\begin{equation}
\omega(|f|^2)=\sum_{IJ} \overline{c}_I c_J (M^{H}\circ M^\perp)_{IJ}
\end{equation}
where the product on the right-hand side is again the Hadamard product, and the matrices 
\begin{equation}
M^H_{IJ}=\mu(T_{\ch{n}^H_J-\ch{n}^H_I}), \qquad M^\perp_{IJ}=\omega_{\text{AIL}}(T_{\ch{n}^\perp_J-\ch{n}^\perp_I})
\end{equation}
are positive semidefinite, because $\mu$ and $\omega_{\text{AIL}}$ are. 

Thus for each puncture data $\mathcal{P}$, we have found a state on the charge networks. The resulting GNS representation has the properties we have discussed in the previous section on $H$, and the standard properties of the AL representation away from $H$. We note that according to general results \cite{Ashtekar:1994mh}, the state $\omega$ defines a measure on generalized connections on $\spat$.  

Given this representation of ``half'' of the holonomy-flux algebra $\mathfrak{A}$, we will now briefly discuss whether the flux operators $E_S$ can be represented alongside the charge networks. How this question can be answered for general representations has been discussed in \cite{Sahlmann:2002xu,Sahlmann:2002xv}. The main obstruction is that the fluxes have to be symmetric, thus for nontrivial measures on generalized connections, a `divergence term' has to be added to the derivative, see \eqref{eq:div}. This divergence term may fail to exist in the proper sense, thus making it impossible to represent flux operators. In the present case, the situation is as follows: 
\begin{itemize}
	\item All flux operators $E_S$ with $S\cap H=\emptyset$ can be defined. 
  \item All flux operators $E_S$ with $S\subset H$ can be defined. 
  \item All flux operators $E_S$ with $S\cap H$ a 1-cycle\footnote{The orientation of $S\cap H$ can be defined using the orientation of $S$, the orientation of $\spat$ and the orientation of $H$.} can be defined.
\end{itemize}
The first point is obvious. The second point is due to the fact that the derivations $X_S$ related to flux operators for surfaces $S\subset H$ only act nontrivially on holonomies that intersect $S$, and hence $H$, transversally. But the measure relevant for those edges is just the AL measure, and hence no divergence term is needed. 

There is a problem representing fluxes through $S$ where $S\cap H$ is not a cycle. The problem is related to the fact that the action of the corresponding derivation $X_S$ can turn a function that is null with respect to $\omega$ into one that is not. Consider for example the two nontrivial cycles $\alpha_1$ and $\alpha_2$ in Fig.\ \ref{fi:bad_flux}:
\begin{figure}
	\centerline{\epsfig{file=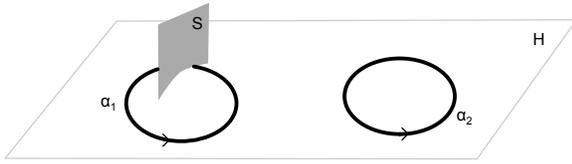,scale=.6}}
	\caption{A surface the flux of which does not have a well defined action on surface states}
	\label{fi:bad_flux}
\end{figure} 
We have 
\begin{equation}
\ket{\alpha_1,n}_{\mathcal{P}}=\ket{\alpha_2,n}_{\mathcal{P}},
\end{equation}
but
\begin{equation}
X_S(T_{\alpha_1,n}-T_{\alpha_2,n})=n T_{\alpha_1,n}-T_{\alpha_2,n},
\end{equation}
and the corresponding GNS vector is \emph{not} zero. For $S\cap H$ a cycle, this cannot happen, and the flux $E_S$ is well defined as we will show in the following. We consider a spin net $T_{\gamma,\ch{n}}$ in $H$. According to \eqref{eq:su1}, it is an eigenstate of $X_S$, 
\begin{equation}
X_S\, T_{\gamma,\ch{n}}=m(S\cap H,\ch{n}) T_{\gamma,\ch{n}}.
\end{equation}
 Take first the case that $\partial \ch{n}=0$. 
Then we claim $m$ is a topological invariant of $S\cap H,\ch{n}$. To see this, we recall that the intersection of two 1-cycles in $H$ defines a 0-cycle, and its homology class depends only on the homology classes of the two 1-cycles, by means of Poincar\'e duality. Now $H_0(H,\Z)=\Z$, so this class is given by an integer. We apply this to the case of the cycles $S\cap H$ and $\ch{n}$. 
The integer is then precisely given by $m(S\cap H,\ch{n})$ up to a global sign that depends on conventions. That this is so can be seen by inspecting the way the coefficients work in \eqref{eq:su1} and in the definition of the intersection 0-cycle. Thus $m(S\cap H,\ch{n})$ is a topological invariant,
 and the definition 
\begin{equation}
E_S\ket{\ch{n}}_\mathcal{P}=m(S\cap H,\ch{n})\ket{\ch{n}}_\mathcal{P}
\label{eq:fluxred}
\end{equation}
makes sense, and is manifestly symmetric. Note that $m$ is zero if $\ch{n}$ is a trivial cycle with respect to the homology of $H$ (i.e. contractible, possibly by crossing punctures), so in the cases in which symmetry of the operator might be an issue, namely when it acts on a loop around a puncture, the action is trivial.  

The case where $\partial\ch{n}\neq 0$ can be treated with a similar argument, by noting that for two charge nets $\ch{n}$ and $\ch{n}'$, 
\begin{equation}
\ket{\ch{n}}_\mathcal{P}=\text{const. } \ket{\ch{n}'}_\mathcal{P}
\end{equation}
precisely when they are in the same homology class in $H_1(H,\Z)$. But in that case, as remarked above, they differ by a boundary $\partial S'$, and 
\begin{equation}
\begin{split}
m(S\cap H,\ch{n})T_{\ch{n}}&=X_S[T_{\ch{n}}]
=X_S [T_{\ch{n}'}T_{\partial S'}]\\
&= X_S[T_{\ch{n}'}] T_{\partial S'}+T_{\ch{n}'} X_S[T_{\partial S'}]\\
&=X_S[T_{\ch{n}'}] T_{\partial S'}\\
&=m(S\cap H,\ch{n}')T_{\ch{n}},
\end{split}
\end{equation}
whence $m(S\cap H,\ch{n})=m(S\cap H,\ch{n}')$ and \eqref{eq:fluxred} is well defined also in this case. 

Thus we have a representation of a large class of fluxes, in fact, as many as we can expect to recover, given that \eqref{eq:hu1} freezes many of the degrees of freedom on $H$.  

\subsection{Gauge invariant solutions}
It remains to write down gauge invariant solutions to \eqref{eq:hu1}. To this end, consider puncture data 
\begin{equation}
\mathcal{P}= \{(p_1,m_1),(p_2,m_2), \ldots, (p_N,m_N)\}
\end{equation}
and work in the GNS representation corresponding to this data. 
Then it is easy to see the following: The space of solutions is spanned by charge networks $\ch{n}$ that 
\begin{enumerate}
	\item intersect $H$ precisely in the punctures $p_1
	\ldots p_N,$
	\item satisfy, for all punctures $p_i$,
	\begin{equation}
\sum_{e \text{ at } p_i} \sigma(e) n_e = m_i, 
\end{equation}
\end{enumerate}
We note that these solutions are not necessarily gauge invariant. The gauge invariant solutions form a subspace. 

Let us make a few comments on the structure of the representation with puncture data $\mathcal{P}$ and the space of gauge invariant solutions. 

\begin{figure}
	\centerline{\epsfig{file=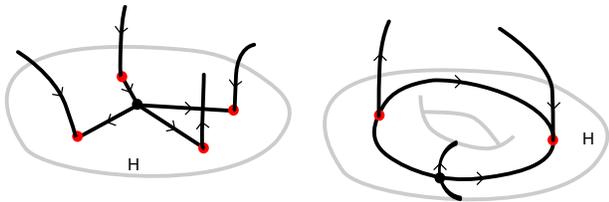,scale=.6}}
	\caption{Graphs of two gauge invariant charge nets, for the case of $H\simeq S^2$ (left) and $H\simeq T^2$ (right).}
	\label{fi:gauge_inv}
\end{figure}

\emph{The case $H\simeq S^2$. } The case in which $H$ has $S^2$ topology is the one usually considered for a black hole horizon. Then there are no nontrivial cycles. Consequently, there is no nontrivial gauge invariant charge network that is lying entirely in $H$, 
\begin{equation}
\ket{\ch{n}}_\mathcal{P}=\text{const.} \ket{0}_\mathcal{P} \quad \text{ for } \quad \ch{n}\subset H.
\end{equation}
For a general gauge invariant charge network $\ch{n}$, we find that due to the diffeomorphism invariance of the measure on $\abar_H$, we can always label the 
part $\ch{n}^H$ that is lying in $H$ in such a way that there is at most one edge emanating from each point on $H$ that is intersected by $\ch{n}^\perp$, and all of these edges meet in a single point (see the left-hand part of Fig.\ \ref{fi:gauge_inv}).  
Furthermore, due to the fact that in this case a loop surrounding all the punctures is contractible in $H'$, we get \eqref{eq:zeromodk}.

\begin{figure}
	\centerline{\epsfig{file=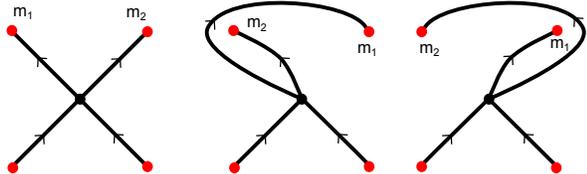,scale=.6}}
	\caption{Two ways to permute two punctures: Left, the original configuration, middle and right, the two final states of the permutation.}
	\label{fi:transposition}
\end{figure}

Finally, so far we have only studied the action of diffeomorphisms that keep the punctures fixed. Let us briefly take a glimpse at the action of diffeomorphisms that move punctures (those transformations will move states between different GNS Hilbert spaces). One would eventually like to mod out these transformations, but this is a nontrivial task due to the following fact: When one considers the process of exchanging two punctures by diffeomorphisms that leave the other punctures invariant, one finds that there are two ways to do it, see Fig.\ \ref{fi:transposition}. 
There must be a relative phase that is picked up on the way, because the two final states differ by a phase
\begin{equation}
\varphi=e^{\frac{2\pi i}{k}(m_2n_2-m_1n_1)}
\end{equation}
This means that the punctures must obey some kind of anyonic statistics. This is very encouraging, as it was found when studying the entropy of isolated horizons, that  punctures can not behave like identical particles. They must be, to a certain extent, distinguishable to account for a linear area-entropy relation \cite{Domagala:2004jt}. 

\begin{figure}
	\centerline{\epsfig{file=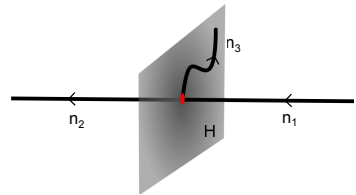,scale=.6}}
	\caption{Gauge invariance gives a condition on $n_1,n_2$, and $n_3$, but does not fix the flux $m=-(n_1+n_2)$ (`outside' of $H$ is to the right)}
	\label{fi:gauge_flux}
\end{figure}

We note that in all of these aspects, the U(1) case compares very well to the 
quantization of an isolated horizon when gauge fixing to $U(1)$ before quantization. Here too, the state on $H$ of solutions of the horizon condition 
is uniquely determined by the structure of the punctures. The connection on $H$ can be considered flat except at the punctures, and the punctures obey nontrivial statistics. The fluxes $m_1\ldots m_N$ determine the measure only $\mod k$, and their sum must be 0 $\mod k$ for consistency. The only difference is that due to the fact that $H$ does not have to be a boundary of $\spat$, there is no strict correlation between the flux through $H$ at a puncture, and the flux of gauge charge into $H$. Consider for example the puncture in Fig.\ \ref{fi:gauge_flux}:
The measure on $\abar_H$ near the puncture is fully determined by the flux
$m=-(n_1+n_2)$ through $H$ at the puncture. Gauge invariance gives the condition $n_1=n_2+n_3$. Thus there are different configurations with the same flux possible at a given puncture.     

\begin{figure}
	\centerline{\epsfig{file=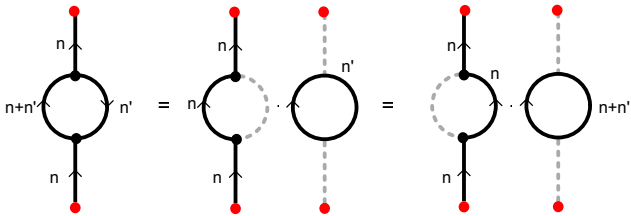,scale=.6}}
	\caption{Change in decomposition of a $H$-charge net}
	\label{fi:tree_change}
\end{figure}

\emph{The general case}: The case where $H$ has more general topology shows all the same features that we have described above for $H\simeq S^2$. The only difference is that there is a nontrivial space of gauge invariant charge network states that lie entirely within $H$. For genus $g$ there are $2g$ nontrivial cycles which contribute nontrivial holonomies (see Fig.\ \ref{fi:gauge_inv}, right-hand side). In this case, for a gauge invariant charge net $\ch{n}$, the part $\ch{n}_H$ in $H$ can be decomposed into a part connected with the punctures, with a single internal vertex analogous to what happened in the $S^2$ case and a part completely internal to $H$. This decomposition is however not necessarily unique, see Fig.\ \ref{fi:tree_change}.
\section{Towards quantum isolated horizons with SU(2) boundary conditions}
\label{se:su2}
Now we will describe, how far we can get for the SU(2) case, in taking the same steps as in the U(1) model. The goal is again to construct state $\omega$ that induce GNS representation of the holonomy-flux algebra containing solutions to \eqref{eq:q_horizon}.

The crucial difference between the U(1) model investigated so far, and the situation for SU(2) is that due to the non-Abelian nature of SU(2), the operators $\op{W}_S$ (see \eqref{eq:exph}) are highly nontrivial. In the U(1) case, the corresponding object was just the exponential of an operator, \eqref{eq:hu1}. Thus let us start with a summary of the properties of these operators that can be gleaned from \cite{Sahlmann:2011uh}. 
\subsection{The surface operators $\op{W}_S$}
\label{se:ws}
In this section we will discuss the properties of the operators $\op{W}_S$. 
All the results we give are either contained in \cite{Sahlmann:2011uh} or easily obtainable with the methods contained in that work. A more thorough investigation of the properties of  the $\op{W}_S$ is still desirable, and will be undertaken elsewhere \cite{sahl1}.  

The first property we want to list is that given two surfaces $S_1$, $S_2$ that 
intersect each other at most nontransversally, i.e., in such a way that they are both contained in a bigger surface $S$, then the traces of the corresponding operators commute, 
\begin{equation}
[\tr_j(\op{W}_{S_1}), \tr_{j'}(\op{W}_{S_2})]=0, \qquad \text{for } S_1,S_2\subset S. 
\end{equation}
Here and in the following, $\tr_j$ stands for the character in the irreducible representation of SU(2) labeled by the half-integer $j$, 
\begin{equation}
\tr_j(\op{W}_{S})=\tr(\pi_j(\op{W}_{S})).
\end{equation}
The next property that will be important for us is that traces of the operator $\op{W}_S$ are diagonalized by spin network edges that pierce the surface $S$. For states $\Psi_j$ having one spin-$j$ edge puncturing $S$ and exiting $S$ on the other side, we find
\begin{equation}
\begin{split}
\tr_{\frac{1}{2}}(\op{W}_S)\Psi_{0}&=\frac{q-q^{-1}}{q^{\frac{1}{2}}-q^{-\frac{1}{2}}}\,\Psi_0,\\
\tr_{\frac{1}{2}}(\op{W}_S)\Psi_{\frac{1}{2}}&=(q+q^{-1})\Psi_{\frac{1}{2}}
\end{split}
\label{eq:ev1}
\end{equation}
and one can show that higher spin punctures continue to be eigenvectors for $\tr_{\frac{1}{2}}{W}_S$. The constant $q$ in the above formulas is given by 
\begin{equation}
q=\exp\left(\frac{2\pi i}{k}\right),
\end{equation}
with 
\begin{equation}
k=-\frac{2a_H}{\pi\beta(1-\beta^2)\lp^2}.
\end{equation}
Furthermore, one can show that this pattern continues for traces in other irreducible representations, 
\begin{equation}
\tr_{j}(\op{W}_S)\Psi_{j'}=c_{j,j'}\Psi_{j'},
\label{eq:ev2}
\end{equation}
but we have no closed formula for the eigenvalues $c_{j,j'}$. We note that 
$c_{1/2,0}$ is the path integral expectation value for an unknotted Wilson loop in the $j=1/2$ representation, in SU(2) Chern-Simons theory with level $k$, and $c_{1/2,1/2}$ is related to the expectation value of linked Wilson loops \cite{Sahlmann:2011uh}. We thus suspect that all the $c_{j,j'}$ are related to SU(2) Chern-Simons theory in a similar way. 

Moreover, one can see that states $\Psi'_j$ having one spin-$j$ edge puncturing $S$ and \emph{ending} on $S$ are again eigenvectors, 
\begin{equation}
\tr_{j}(\op{W}_S)\Psi'_{j'}=c'_{j,j'}\Psi'_{j'}.
\label{eq:ev3}
\end{equation}
We stress  that the eigenvalues in \eqref{eq:ev1}, \eqref{eq:ev2},\eqref{eq:ev3} are completely independent of the shape of $S$, as long as the boundary of $S$ encloses the puncture. (This is obvious for \eqref{eq:ev1}, but it is also true in all the other cases, as can be seen with the methods in \cite{Sahlmann:2011uh}.)

The cases that several edges pierce $H$ in the same puncture, and that a surface contains several punctures are more complicated, and there are indications that the spin nets puncturing the surface may not be eigenstates in general. 

It is \emph{not} true that the operators $\op{W}_S$ \emph{themselves} are diagonalized by the states $\Psi_{j}, \Psi'_{j}$. What happens is that the surface $S$ in \eqref{eq:exph} comes with extra structure from the non-Abelian Stokes Theorem: This involves the choice of a system of paths in $\mathcal{S}$, connecting the beginning/endpoint of $\alpha$  with the points of $S$. When acting on a state $\Psi_{j}$, the operator will in general give back a sum of spin networks, some of which involve holonomies along the path system in $S$ that are coupled via suitable intertwiners to the spin network edge piercing the surface $S$. There is, however, a crucial exception: Going through the same steps that were used in \cite{Sahlmann:2011uh} to calculate 
$\tr_{1/2}(\op{W}_S)\Psi_{0}$, one finds 
\begin{equation}
\op{W}_S\Psi_{0}= c\um_{2}\Psi_{0}.
\label{eq:important}
\end{equation}
This result is natural, since for the state $\Psi_{0}$, there is nothing the holonomies along the path system in $S$ can couple to. It should however also be said that the case $\Psi_{0}$ is special in that there is a divergence in the eigenvalue that has to be renormalized away \cite{Sahlmann:2011uh}, so this case merits further careful investigation. In particular, it is not entirely clear what the value of finite constant $c$ leftover in \eqref{eq:important} should be. If we follow the argument in \cite{Sahlmann:2011uh}, the answer would be 
\begin{equation}
c=\frac{1}{2}(q^{\frac{1}{2}}+q^{-\frac{1}{2}}). 
\end{equation}
It is noteworthy that in \cite{Sahlmann:2011uh}, $c$ sets the normalization of the Jones polynomial, which in principle is arbitrary. Thus it is conceivable that another way to remove the divergence would yield another value, in particular, $c=1$.  

A final remark is about the properties of $\op{W}_S$ under change of orientation of $S$. From \cite{Sahlmann:2011uh} we know
\begin{equation}
\tr_j(\op{W}_{-S})=\tr_j(\op{W})^\dagger_S,
\label{eq:dagger}
\end{equation}
where $-S$ is obtained from $S$ by change of orientation. 
\subsection{A functional on simple loops}
\label{se:flatsu2}
Now we can come back to our main topic: What do the properties of $\op{W}_S$ listed above mean for our goal of defining a measure on $\abar_H$ through \eqref{eq:exph}? The first thing to note is that because $\tr_{j}(\op{W}_S)$ 
are diagonal on the states $\Psi_{j}, \Psi'_{j}$, so must be the traces $\tr_{j}(h_\alpha)$ of holonomies around loops $\alpha$ in $H$. In particular their quantum-mechanical fluctuations must vanish, 
\begin{equation}
\expec{(\tr_{j}(h_\alpha))^2}_\mathcal{P}- \expec{\tr_{j}(h_\alpha)}_\mathcal{P}^2=0,
\end{equation}
which shows that the gauge invariant information contained in contractible loops on $H$ is completely fixed by the puncture data $\mathcal{P}$, and supports the idea that representations that contain horizons should be based on a measure of a form similar to \eqref{eq:meas}. 

To formalize this, let us again pick puncture data
\begin{equation}
\mathcal{P}=\{(p_1,j_1,m_1),(p_2,j_2,m_2),\ldots (p_N,j_N,m_N)\},
\end{equation}
where $p_1 \ldots p_N$ are points on $H$ and $j_1 \ldots j_N$ and $m_1 \ldots m_N$ are labels of irreducible representations of SU(2), and magnetic quantum numbers in those representations.  Furthermore, We will denote traces of holonomy functionals as  
\begin{equation}
T_{\alpha,j}[A]:= \tr_j(h_\alpha[A]).
\end{equation} 
Now let $\{\alpha_i\}$ be a collection of loops that are contractible within $H$, and such that each one of them encloses at most one puncture in $\mathcal{P}$. We will also call such loops \emph{simple}. Then to each simple loop there is an oriented disc $S_i$ in $H$, such that $\partial S_i=\alpha_i$, and we can set 
\begin{equation}
\mu(\prod_i T_{\alpha_i,k_i}):= \prod_i c'_{k_i,j_i}
\label{eq:su2def}
\end{equation}
where the numbers $c'$ are the eigenvalues from \eqref{eq:ev2}, i.e., we have assumed that the punctures $\mathcal{P}$ come from edges ending in $H$. This
can be generalized in an obvious way to the case where some of the piercing edges do not end on $H$. Linear extension of this definition gives a functional on a large class of gauge invariant functionals of the pullback of $A$ to $H$.

Next we check positivity: The first remark is that due to the unitarity of the irreducible representations of SU(2), we have 
\begin{equation}
\overline{T_{\alpha,j}[A]}=T_{-\alpha,j}[A], 
\end{equation}
where we denote with $-\alpha$ the change of orientation of $\alpha$. 
If $\alpha=\partial S$, then $-\alpha=\partial(-S)$. Taking this together with 
\eqref{eq:dagger} gives 
\begin{equation}
\mu(\overline{T_{\alpha,j}})=\overline{\mu(T_{\alpha,j})}.
\end{equation}
Similar statements can be made for products of traces, due to the factorization property of \eqref{eq:su2def}. With this said, let $\{\ch{\alpha}_I\}$ be a collection of  multiloops, where the individual loops again fulfill the requirements assumed in the definition of $\mu$, and 
$\{\ch{k}_I\}$ corresponding assignments of representations. Then we consider 
the expectation value of the modulus squared of the function 
$f=\sum_I c_I T_{\ch{\alpha}_I, \ch{k}_I}$,   
\begin{equation}
\begin{split}
\mu(|f|^2)&=\sum_{IJ} \overline{c_I}c_J \mu(\overline{T_{\ch{\alpha}_I,\ch{k}_I}}T_{\ch{\alpha}_J,\ch{k}_J})\\
&=\sum_{IJ} \overline{c_I}c_J\mu(\overline{T_{\ch{\alpha}_I,\ch{k}_I}})
\mu(T_{\ch{\alpha}_J,\ch{k}_J})\\
&=\sum_{IJ} \overline{c_I}c_J\overline{\mu({T_{\ch{\alpha}_I,\ch{k}_I}})}
\mu(T_{\ch{\alpha}_J,\ch{k}_J})\\
&\geq 0,
\end{split}
\end{equation}
so the functional is positive. This is very encouraging. There is, however, another important property that we can not yet check for $\mu$. Because of the properties of the operators $\op{W}_S$, we have defined $\mu$ such that it factorizes, 
\begin{equation}
\mu(T_{\alpha,k}T_{\alpha',k'})=\mu(T_{\alpha,k})\mu(T_{\alpha',k'}).
\end{equation}
The problem is that the product among the functionals $T_{\alpha,j}$ is not free. For example, we have  
\begin{equation}
T_{\alpha,k}T_{\alpha,k'}=\sum_{k''} c_{k''} T_{\alpha,k''}
\end{equation}
for some constants $c_{k''}$.
We have no formal proof that $\mu$ is compatible with these relations, but we have some indication that that is indeed the case. The point is that the expression 
\begin{equation}
\ppexp{S} \mathcal{F}[A]\,\text{d}^2 s
\end{equation}
that is identical to a holonomy by virtue of the non-Abelian Stokes theorem differs from the operator $\op{W}_S$ only by the fact that certain polynomials in $F$ are replaced by operators in the universal enveloping algebra U(SU(2)) of SU(2). Therefor, $\op{W}_S$ can lose properties of a holonomy only at that point. Now, the replacement of polynomials of $F$ by operators is done using the Duflo map, which insures that the replacement is an isomorphism of algebras on the gauge invariant polynomials of $F$, thus on this subspace it does not lose any structure. Moreover, in the calculations of eigenvalues in \cite{Sahlmann:2011uh}, only that subspace was relevant. Thus we conjecture 
\begin{equation}
c'_{k,j}c'_{k',j}=\sum_{k''} c_{k''} c_{k'',j},
\end{equation}
and similar relations among the eigenvalues of the $\op{W}_{S}$ that make $\mu$ consistent. 

If we grant consistency of $\mu$ as above, what is left to do? The algebra of the simple loops above does not include holonomies around noncontractible loops, nor does it contain gauge noninvariant functionals. We have to show that the functional extends consistently to this larger class of holonomy functionals on $H$. Then it can be extended further to $\cyl$ with exactly the same arguments used in Sec.\ \ref{se:ext}. Finally one would have to consider the action of the fluxes, but the situation is exactly the same as for the fluxes in the U(1) case, and so we forsee no difficulty with defining all the physically relevant flux operators. Since the bookkeeping involved in these steps is quite complicated, we leave their completion to another work \cite{sahl2}. We will finish by commenting on some ramifications of the picture that emerges.  

Is the state $\mu$ extended to all cylindrical functions, supported on connections that are locally flat, except at the punctures? The answer seems to be yes. Note first that the definition of the state $\mu$ on simple loops \eqref{eq:su2def} makes no reference to the precise location of the loops, apart from that they must enclose at most one puncture. Note furthermore that due to \eqref{eq:important}, for $\alpha\in H$ the operator $h_\alpha$ (not just its trace) must be represented by a multiple of the identity due to the measure \eqref{eq:meas}. This makes holonomies $h_e$ in $H$ dependent only on the homotopy class of $e$ in $H$ with the punctures removed.
\begin{figure}
	\centerline{\epsfig{file=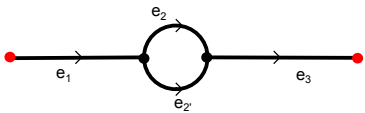,scale=.6}}
	\caption{The graph used in the discussion of diffeomorphism invariance in the SU(2) case}
	\label{fi:tree_change_na}
\end{figure}
For example, consider $h_e$ for the path $e=e_1\circ e_2 \circ e_3$ depicted in Fig.\ \ref{fi:tree_change_na}. We have
\begin{equation}
\begin{split}
h_{e_1}h_{e_2}h_{e_3}&=h_{e_1}h_{e_2}h_{e'_2}h_{e'_2}^{-1}h_{e_3}\\
&=h_{e_1}h_\alpha h_{e'_2}^{-1}h_{e_3}\\
&=c h_{e_1}h_{e'_2}h_{e_3}.
\end{split}
\end{equation} 
Certainly this is acceptable only if $c=1$, but as we have said above, one can argue that $c$ depends on normalization and has to be set by hand, anyway. If this is done, we have indeed  $h_e\Psi=h_{e'}\Psi$ for the path $e'=e_1\circ e'_2 \circ e_3$ and $\Psi$ a solution, and similar identities in more complicated situations.  

Thus altogether, given what we know about the operators $\op{W}_S$, it seems that the program of finding rigorous solutions to \eqref{eq:exph} has a chance to succeed, in a very similar way as it did for the case of the U(1) theory. Obviously, there are still some steps to be taken. There is work in progress on these issues, and results will be reported in a future publication \cite{sahl2}.  
\section{Discussion and outlook}
\label{se:disc}
\begin{figure}
	\centerline{\epsfig{file=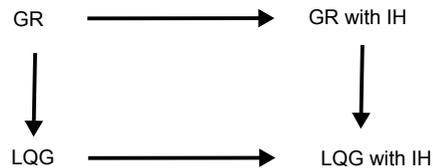,scale=.6}}
	\caption{Two ways to obtain a quantized isolated horizon}
	\label{fi:commut}
\end{figure}

In the present work we have taken the condition \eqref{eq:horizon} for an isolated horizon from classical general relativity, expressed it as an equation [\eqref{eq:hu1}, and \eqref{eq:exph}, respectively] in the quantum theory using operators in the holonomy-flux algebra $\mathfrak{A}$ of \lqg, and studied one class of solutions, both for a U(1) model and for the full SU(2) theory. In the U(1) case, we could complete all the steps, in the SU(2) case there are still some open questions. Our procedure is quite different from the one followed so far \cite{Ashtekar:1997yu,Ashtekar:2000eq,Engle:2009vc,Engle:2010kt} in which one, roughly speaking, quantizes the phase space of space-times that contain an isolated horizon as an inner boundary, see Fig.\ \ref{fi:commut}. 
Classically, the degrees of freedom that live on the horizon are part of the full field content of general relativity. In our treatment, this is transparent in the quantum theory:  The horizon degrees of freedom are represented simply by elements (or components of elements) of the algebra $\mathfrak{A}$ that characterizes \lqg and makes no reference to horizons. 

It was not clear \textit{a priori}, whether the results of the two routes to quantum isolated horizons would coincide, but it turned out that it appears they do to a remarkable extent. In the U(1) case, we recover almost verbatim the structure of the quantized horizon of \cite{Ashtekar:1997yu,Ashtekar:2000eq}. In the SU(2) case, our results are not complete, but what we saw points towards a very similar picture as \cite{Engle:2009vc,Engle:2010kt}. There are differences, however, and they become more obvious if one considers the case of nontrivial horizon topology. Then the picture we obtain on the horizon is not so much resembling quantized U(1) or SU(2) Chern-Simons theory, as quantized  U(1) or SU(2) BF-theory, i.e. Chern-Simons theory with `twice as many degrees of freedom'.  For example, in the torus case we see two holonomies and two conjugate fluxes, whereas one would expect that the two holonomies are conjugate to each other for  U(1) or SU(2) Chern-Simons theory. This does not make much difference for the horizon of spherical topology that is most relevant for the description of black holes, but it may be interesting from a conceptual viewpoint: We find that \emph{\lqg, restricted to certain special null-hypersurfaces, gives Euclidean quantum gravity in three dimensions}. 

From the conceptual standpoint taken in this work, the SU(2) version \eqref{eq:horizon} of the isolated horizon boundary condition seems more attractive, since less classical structure is needed in the quantum theory, and we have a chance to account for all quantum degrees of freedom with the algebra $\mathfrak{A}$. If only one of the three components of the SU(2) connection $A$ is fixed by the horizon conditions, there are two other components free on the horizon. Since they do not form a connection, they are hard to treat with \lqg methods. This is the reason for treating the U(1) case with gauge fixing to U(1) \emph{everywhere}. But if one wishes, one can certainly gauge fix only on the horizon, and our methods will essentially give back the picture of \cite{Ashtekar:1997yu,Ashtekar:2000eq}.

Clearly, we have to complete the investigation of the SU(2) case, but besides and beyond this, there are other interesting questions that should be considered. We make a list of some of them:  
\begin{enumerate}
\item For the case that the horizon is not a boundary of $\spat$, the counting of horizon states of a given area may be modified due to the fact that gauge charge can now enter \emph{and} leave at the punctures. This means that the area-entropy relation may be modified - a potential problem for the approach. This needs to be studied in detail.  

\item It is odd that the classical horizon area $a_H$ shows up in the quantum horizon condition \eqref{eq:q_horizon}, and consequently all over the place in the quantum theory. It would be very nice if this can be changed, either through a change in the classical theory as sketched in \cite{Engle:2010kt} or through replacing  $a_H$ in \eqref{eq:q_horizon} by a suitable operator. 

\item Since the states containing a horizon and the vacuum state of \lqg are states on the same algebra, it is now possible to relate them. In particular, it should be possible to approximate the states with a horizon by states in the vacuum (AL) representation.  

\item The implementation of diffeomorphisms that move punctures should be studied carefully, to confirm the anyonic statistics of the punctures in both, the U(1) and the SU(2) case. 

\item The type of solutions of the quantum horizon condition \eqref{eq:q_horizon} we found may not be the only one. Note for example that we took the flux as given from the vacuum representation of \lqg, and thereby fixed the connection on the horizon. One could think of doing it the other way around, taking the holonomy operators on the horizon to be in the \lqg vaccum, and thereby determine a nonstandard representation of the fluxes. There may be other possibilities. This merits further thought. 

\item Horizons with nontrivial topologies and their quantization in \lqg have been studied in a very interesting series of articles \cite{Kloster:2007cb,Brannlund:2008iw,DeBenedictis:2011hh}. This gives a good point of comparison for the results reported in the present work. 

\end{enumerate}

Finally, this investigation may be the motivation to find and study other kinds of branelike states in \lqg.  
\begin{acknowledgments}
I thank the organizers of the 2011 Shanghai Asia-Pacific School and Workshop on Gravitation, where the idea for this article was first conceived. Madhavan Varadarajan, Lee Smolin, and  Thomas Thiemann gave valuable comments on an earlier version of the article. 

This research was partially supported by the Spanish MICINN Project No.\ FIS2008-06078-C03-03. 
\end{acknowledgments}


\end{document}